\documentclass[times,doublespace]{qjrms4}

\usepackage[colorlinks,bookmarksopen,bookmarksnumbered,citecolor=red,urlcolor=red]{hyperref}

\newcommand\BibTeX{{\rmfamily B\kern-.05em \textsc{i\kern-.025em b}\kern-.08em
T\kern-.1667em\lower.7ex\hbox{E}\kern-.125emX}}

\usepackage{moreverb}
\usepackage{amsmath}
\usepackage{gensymb}
\usepackage{color}
\usepackage{soul}

\begin{document}
\runningheads{B.~L.~Harris and R. Tailleux}{Moist Available Potential Energy Algorithms}

\title{Assessment of Algorithms for Computing Moist Available Potential Energy}

\author{Bethan L.~Harris\corrauth, R\'{e}mi Tailleux}

\address{Department of Meteorology, University of Reading, UK}

\corraddr{B.L. Harris, Department of Meteorology, University of Reading, Earley Gate, PO Box 243, Reading RG6 6BB, UK. E-mail: b.l.harris@pgr.reading.ac.uk}

\begin{abstract}
Atmospheric moist available potential energy (MAPE) has been traditionally defined as the potential energy of a moist atmosphere relative to that of the adiabatically sorted reference state defining a global potential energy minimum. Finding such a reference state was recently shown to be a linear assignment problem, and therefore exactly solvable. However, this is computationally extremely expensive, so there has been much interest in developing heuristic methods for computing MAPE in practice. Comparisons of the accuracy of such approximate algorithms have so far been limited to a small number of test cases; this work provides an assessment of the algorithms' performance across a wide range of atmospheric soundings, in two different locations. We determine that the divide-and-conquer algorithm is the best suited to practical application, but suffers from the previously overlooked shortcoming that it can produce a reference state with higher potential energy than the actual state, resulting in a negative value of MAPE. Additionally, we show that it is possible to construct an algorithm exploiting a theoretical expression linking MAPE to Convective Available Potential Energy (CAPE) previously derived by Kerry Emanuel. This approach has a similar accuracy to existing approximate sorting algorithms, whilst providing greater insight into the physical source of MAPE. In light of these results, we discuss how to make progress towards constructing a satisfactory moist APE theory for the atmosphere. We also outline a method for vectorising the adiabatic lifting of moist air parcels, which increases the computational efficiency of algorithms for calculating MAPE, and could be used for other applications such as convection schemes.
\end{abstract}

\maketitle
\section{Introduction}
\label{sec:intro}
Available Potential Energy (APE) theory, as originally outlined by \citet{lorenz1955available}, provides a framework to study the energy available to atmospheric motions. The theory is underpinned by the concept of an atmospheric background or \textit{reference} state. Such a state has been traditionally envisioned as being obtained through an adiabatic mass rearrangement such that the sum of the internal and potential energies of the atmosphere (total potential energy) is minimised. The APE is then found as the difference between the total potential energy of the atmosphere and the total potential energy of the reference state. In its reference state, the atmosphere is at rest and in hydrostatic equilibrium; its density stratification is therefore statically stable and horizontally uniform, and no further conversion with kinetic energy can take place. The APE thus gives the total potential energy that is available for reversible conversions into kinetic energy. Assuming hydrostatic balance, minimisation of the total potential energy is equivalent to minimisation of the enthalpy $H$, so that
\begin{equation}
\label{eq:APE}
\mathrm{APE} = \int{\left(h - h_{\mathrm{ref}}\right)}\, \mathrm{d}m,
\end{equation}
where $h$ is the specific enthalpy, and the integral is over all the mass in the considered atmospheric domain.

For a moist atmosphere, the rearrangements are made via reversible adiabatic processes conserving total water content \citep{lorenz1978available}. In this work, we refer to the APE of a moist atmosphere as MAPE (Moist Available Potential Energy), following the terminology of \citet{stansifer2017accurate}, and we focus only on the vertical component of MAPE. Unlike the dry case, for which reference pressure is uniquely determined by sorting potential temperature, there is no known analytical solution for obtaining the moist reference state from the distribution of entropy and specific humidity. As a result, previous methods of calculating MAPE have relied on heuristic approaches involving discretising atmospheric domains into parcels of equal mass and sorting them according to density at differing pressure levels to obtain a reference state. From a computational viewpoint, the discretised approach to computing MAPE is equivalent to finding the permutation of the actual state with the lowest total potential energy. \citet{tailleux2004seemingly} characterised such a problem as an asymmetric travelling salesman problem, but recently, it was realised by \citet{hieronymus2015finding} that the computation of such a reference state was in fact a linear assignment problem that can be solved by using the Munkres algorithm \citep{munkres1957algorithms}. Whilst the Munkres algorithm is exact, it is also computationally expensive, and therefore it is still desirable to use approximate algorithms for speed. The time taken for the algorithms to compute MAPE is explored in more detail in Appendix \ref{sec:bisection}. The Munkres algorithm can be used when the considered atmospheric domain comprises a small number of parcels $n$, but the runtime of the algorithm increases as $n^3$ \citep{stansifer2017accurate}, so it quickly becomes infeasible for large domains.

Approximate sorting algorithms have been employed to investigate the intensity of extratropical storm tracks \citep{ogorman2010understanding}, using Lorenz's algorithm \citep{lorenz1979numerical} to calculate MAPE, and the energetics of tropical cyclones \citep{wong2016computation}, using the top-down and bottom-up algorithms. A review of existing approximate sorting algorithms is given by \citet{stansifer2017accurate}, who discussed their accuracy compared to the exact Munkres algorithm. However, the comparison was made over only three test case soundings. This showed that none of the approximate algorithms was able to compute the exact MAPE in every case, but clearly the small number of cases presented means that it is impossible to draw conclusions about the general relative performance of the algorithms, and therefore difficult to know which is most useful to study atmospheric energetics.

It is also not certain that using parcel-sorting algorithms calculates the most physically suitable form of MAPE. Finding the exact minimum enthalpy parcel rearrangement using the Munkres algorithm does not consider whether certain parcel movements may be restricted, for example by the presence of Convective Inhibition (CIN). The bottom-up algorithm introduced by \citet{wong2016computation} is designed to prevent the unrealistic release of Convective Available Potential Energy (CAPE) during sorting, but does not directly consider either CAPE or CIN in its computation.

In Section \ref{sec:algorithms}, we briefly describe all the existing algorithms that have been designed to calculate MAPE. To investigate the possibility of using a more physically-based approach to compute MAPE, we also develop an algorithm based on the relationship between CAPE and MAPE found by \citet{emanuel1994atmospheric}. As far as we are aware, this relationship has never been explored to investigate whether it can be used to obtain similar results to those of the parcel-sorting approaches.

We then apply all the MAPE algorithms to 3130 soundings from the Atmospheric Radiation Measurement (ARM) station on Nauru, and to 584 soundings from the ARM sites on the Southern Great Plains. This allows us to assess which of the approximate algorithms are likely to compute a MAPE close to the true value, and to investigate the variation in their accuracy over a large number of soundings. In Section \ref{sec:data} we describe the data used for the assessment. Section \ref{sec:comparison} presents the results of the approximate algorithms' performance against the Munkres algorithm, and compares their accuracy between the two locations. In Section \ref{sec:discussion} we discuss how the results relate to what was previously known about the algorithms, and which algorithms are most suitable for practical application. We also discuss the implications of our results for the development of a satisfactory theory of APE for a moist atmosphere. Finally, Appendix \ref{sec:bisection} details the bisection method used to decrease the time taken by each algorithm to compute MAPE, which allowed their efficient application to such a large number of soundings. This method works by allowing vectorised computation of the temperatures of parcels when they are lifted reversibly adiabatically; it is generalisable to any application requiring the adiabatic lifting of parcels, such as a convection scheme.

\section{Algorithms for Computing MAPE}
\label{sec:algorithms}

In this section we describe the algorithms that may be used to compute the reference state, and hence the MAPE, of an atmospheric sounding. We assume here that the sounding has been discretised into parcels of equal mass. To begin, we outline the Munkres algorithm, which finds the reference state corresponding to the exact minimum enthalpy rearrangement of the parcels. We then describe the parcel-sorting algorithms that have been designed to find approximations to the reference state. Due to their approximate nature, these methods are less computationally expensive than the Munkres algorithm, but their typical accuracy compared to the Munkres algorithm is unknown; this will be investigated in Section \ref{sec:comparison}. Finally we describe a method for calculating MAPE that does not rely on a sorting procedure, but instead makes use of the relationship between MAPE and CAPE, which was suggested by \citet{emanuel1994atmospheric}.

\subsection{Munkres algorithm}
\label{subsec:munkres}
The Munkres algorithm \citep{munkres1957algorithms} may be used to obtain the exact minimum enthalpy rearrangement of a set of air parcels, by treating the computation of the parcels' reference pressures as a linear assignment problem \citep{hieronymus2015finding,stansifer2017accurate}. This method first calculates a \textit{cost matrix} $C$, in which the entry $c_{ij}$ is the enthalpy of the $i$\textsuperscript{th} parcel at the $j$\textsuperscript{th} pressure level. Using this cost matrix, the algorithm allocates parcels to the pressure levels resulting in a minimised total enthalpy. This is done by using the linear algebra procedure described by \citet{munkres1957algorithms}, which tracks how difficult it is to find a low-enthalpy position for each parcel during the rearrangement process.

\subsection{Lorenz's algorithm}
\label{subsec:lorenz}
The first algorithm for approximating the minimised enthalpy reference state of a moist sounding was developed by \citet{lorenz1979numerical}. For a set of $n$ parcels at pressures $p_1<p_2<\dotsc <p_n$, this algorithm begins by calculating the virtual temperatures of all parcels as if they were lifted reversibly and adiabatically to $p_1$, denoted $T_{v1}$, and if they were similarly lifted to $p_n$, denoted $T_{vn}$. The algorithm first finds a parcel to assign to pressure level $p_1$, and then moves to progressively higher pressures. This assignment is determined as follows: at each level $p_j$, the unassigned parcels with the highest values of $T_{v1}$ and $T_{vn}$ are identified. If both these values are maximised by the same parcel, this parcel is assigned to $p_j$. If the two identified parcels differ, then their virtual temperatures at $\frac{p_j + p_{j+1}}{2}$ are calculated. The parcel with the higher $T_v$ here is assigned to $p_j$. After $n$ assignments are made in this way, all parcels will have been assigned a different reference pressure, thus determining the reference state. Equivalently, the specific volume may be maximised at each pressure rather than the virtual temperature, as has been done in our implementation.

\subsection{Randall and Wang's algorithm}
\label{subsec:randall_and_wang}
\citet{randall1992moist} noted that it was possible for Lorenz's algorithm to return a negative MAPE, and designed a similar algorithm that eliminated this problem. For pressure levels $p_1<p_2<\dotsc <p_n$ as before, the procedure begins by labelling $p_A = p_1,\, p_B = p_n$. Once again, the virtual temperatures for all parcels are calculated as if they were lifted to $p_A$ and $p_B$, and those parcels with the highest values of $T_{vA}$ and $T_{vB}$ are identified. At this point, the two methods diverge. Randall and Wang next compute the total atmospheric enthalpy for two situations: if the parcel with the highest $T_{vA}$ were lifted to $p_A$, with any intermediate parcels shifted down one pressure level; and if the parcel with the highest $T_{vB}$ were lifted to $p_A$ and the intermediate parcels shifted down. Whichever of these configurations results in the lowest total enthalpy is accepted as the new rearrangement, and $p_A$ is redefined as $p_A=p_2$. The method proceeds until $p_A=p_B$.

\subsection{Top-down algorithm}
\label{subsec:top_down}
The top-down algorithm was used to compute reference states in the study of APE in tropical cyclones by \citet{wong2016computation}. The performance of the top-down algorithm was also analysed by \citet{stansifer2017accurate}, who referred to it as the ``greedy algorithm". The top-down algorithm for $n$ parcels proceeds as follows: all $n$ air parcels are moved reversibly adiabatically to $p_1$, the lowest pressure in the sounding. Their densities at this pressure are calculated, and the parcel with the lowest density is assigned to have $p_n$ as its reference pressure. This parcel is then eliminated from sorting. The remaining $n-1$ parcels are moved to $p_2$, and again their densities are calculated, and the least dense parcel assigned to $p_2$. The algorithm continues in this way until all parcels have been assigned to a reference pressure level.

\subsection{Bottom-up algorithm}
\label{subsec:bottom_up}
Bottom-up sorting works similarly to top-down sorting, but the parcels are first moved to the highest pressure $p_n$, assigning the parcel with the highest density to this level, and proceeding to lower pressure levels $p_{n-1},p_{n-2}\dotsc$. Bottom-up sorting was suggested by \citet{wong2016computation} to limit the inclusion of CAPE in the definition of MAPE. This may be desirable in practice since not all the CAPE present in the atmosphere will be released, for example due to the presence of Convective Inhibition (CIN).

\subsection{Divide-and-conquer algorithm}
\label{subsec:divide_and_conquer}
The divide-and-conquer algorithm was introduced by \citet{stansifer2017accurate}. It is similar to top-down or bottom-up sorting, but all the parcels are initially moved to the middle pressure level $p_m$, where $m = \left \lfloor \frac{n+1}{2} \right \rfloor$. The $m$ parcels with the lowest density at this pressure are assigned to the sub-domain $\left [ p_1,p_m \right ]$, and the $n-m$ parcels with the highest density are assigned to $\left [ p_{m+1},p_n \right ]$. The algorithm then acts recursively on the two sub-domains. In the three test cases analysed by \citet{stansifer2017accurate}, the divide-and-conquer algorithm was found to calculate the exact minimum enthalpy reference state, even in situations where other approximate algorithms failed to capture significant proportions of the MAPE. However, since the divide-and-conquer algorithm is not an exact enthalpy minimisation procedure, we cannot expect this sorting method to compute the true MAPE in all situations.

\subsection{Estimation from Convective Available Potential Energy}
\label{subsec:emanuel}
Rather than using a parcel-sorting algorithm to compute the vertical component of MAPE, it is natural to consider its relation to Convective Available Potential Energy (CAPE), since both are measures of the energy available to vertical motion in a sounding. This link was noted by \citet{randall1992moist}, who referred to the vertical component of MAPE as GCAPE (Generalised CAPE), but did not explore the link between CAPE and GCAPE. \citet{tailleux2004seemingly} suggested the existence of a functional relationship between CAPE and MAPE, which could permit the inexpensive computation of MAPE. However, it is still not known how CAPE-based measures of atmospheric energetics compare to the Lorenz MAPE of Eq. (\ref{eq:APE}). Here we outline an algorithm for calculating MAPE using the CAPE-dependent equations of \citet{emanuel1994atmospheric}, which we will compare (in Section \ref{sec:comparison}) to the MAPE computed by the sorting algorithms described above.

\citet{emanuel1994atmospheric} supposes that MAPE is due solely to the presence of CAPE in a thin boundary layer of depth $\Delta p_b$. In this case an approximation to the MAPE is given by
\begin{equation}
\label{eq:emanuel_MAPE}
\mathrm{MAPE} \approx \frac{\Delta p_b}{g}\left ( \mathrm{CAPE}_b - \frac{1}{2}\Delta p_b\overline{p^{\kappa-1}}p_0^{-\kappa}R_d\Delta\theta_v \right ),
\end{equation}
where $\mathrm{CAPE}_b$ is the mean CAPE in the boundary layer, $\kappa = \frac{R_d}{c_{pd}}$, and $\Delta\theta_v$ is the change in the virtual potential temperature between the top of the boundary layer at $p_{b,\mathrm{top}}$ and the boundary layer's level of neutral buoyancy, $p_{\mathrm{LNB}}$. The overbar denotes a $\theta_v$-weighted average from $p_{b,\mathrm{top}}$ to $p_{\mathrm{LNB}}$. The first term of Eq. (\ref{eq:emanuel_MAPE}) corresponds to the release of CAPE when the boundary layer rises upwards to its LNB. The second term accounts for the energy change that occurs as a result of the remaining air parcels descending by $\Delta p_b$. We will henceforth refer to MAPE calculated using Eq. (\ref{eq:emanuel_MAPE}) as the \textit{Emanuel MAPE}.

We compute the Emanuel MAPE by calculating the value of Eq. (\ref{eq:emanuel_MAPE}) for $\Delta p_b$ depths ranging from 0 mb to 150 mb, and selecting the maximum value of MAPE returned by any of these $\Delta p_b$ values. We increment $\Delta p_b$ simply by including the next lowest parcel in the sounding. Theoretically, it would be possible to use smaller increments in $\Delta p_b$, and include fractions of parcels in the boundary layer. We have not done this because the sorting algorithms discussed earlier in this section are only able to rearrange whole parcels, so allowing this CAPE-based algorithm to only lift whole parcels provides a fairer comparison of the MAPE.

To compute the boundary layer CAPE, $\mathrm{CAPE}_b$, we use a parcel  with a value of $\theta$ given by the pressure-weighted mean of $\theta$ in the boundary layer, and $q$ given by the mean $q$ in the boundary layer. The CAPE is then
\begin{equation}
\label{eq:CAPE}
\mathrm{CAPE}_b = \int_{p_{\mathrm{LNB}}}^{p_i} \left ( \alpha_p - \alpha_e \right ) \mathrm{d}p,
\end{equation}
where $\alpha_p$ is the specific volume of the parcel when it is lifted reversibly adiabatically, and $\alpha_e$ is the environmental specific volume. The parcel is lifted from its initial position $p_i$, which we take to be the bottom of the boundary layer (i.e. the surface), to its highest level of neutral buoyancy.

\section{Data} 
\label{sec:data}
\begin{figure*}
\centering
\includegraphics[width=0.45\textwidth]{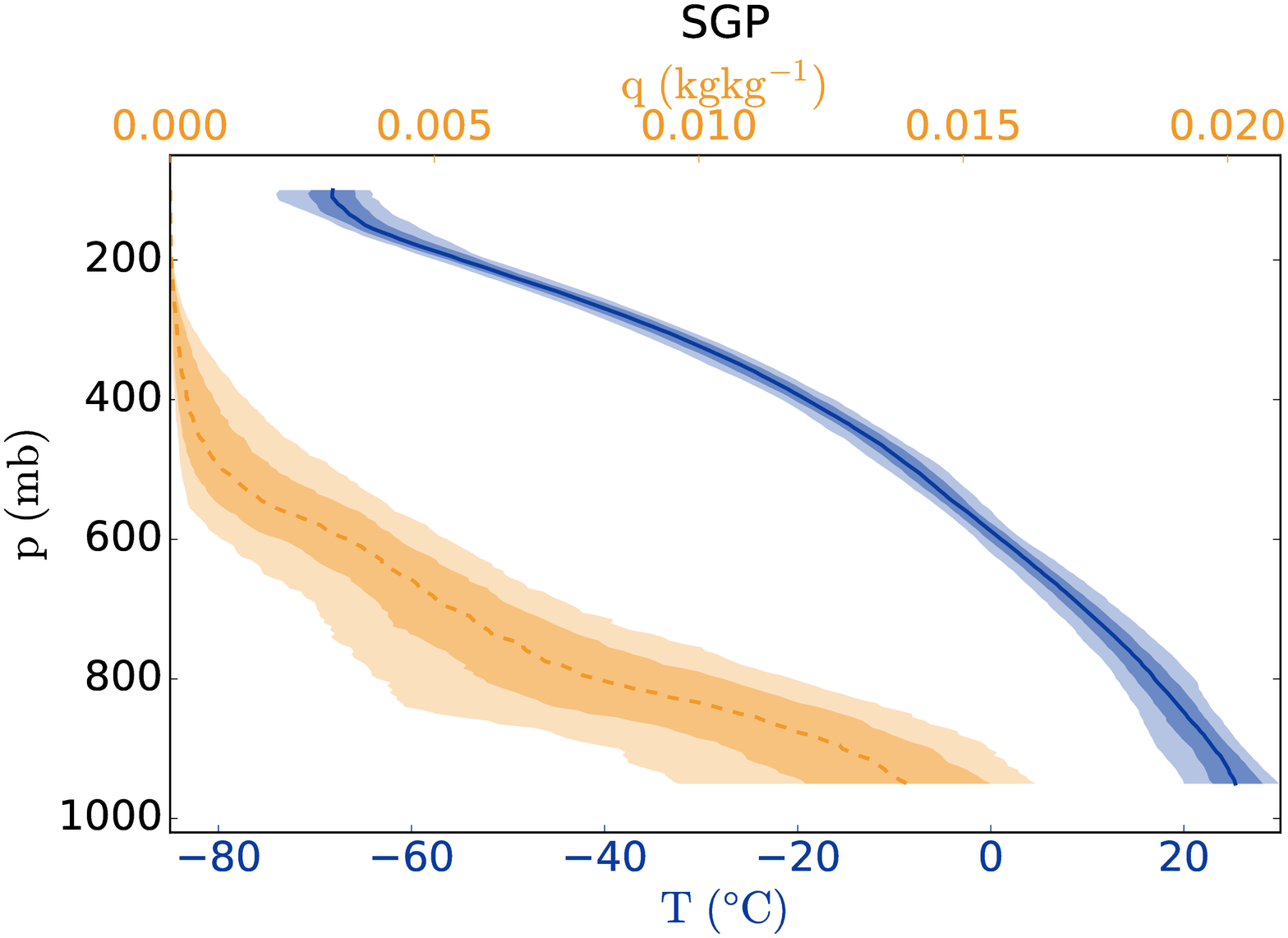}\hspace{3mm}
\includegraphics[width=0.45\textwidth]{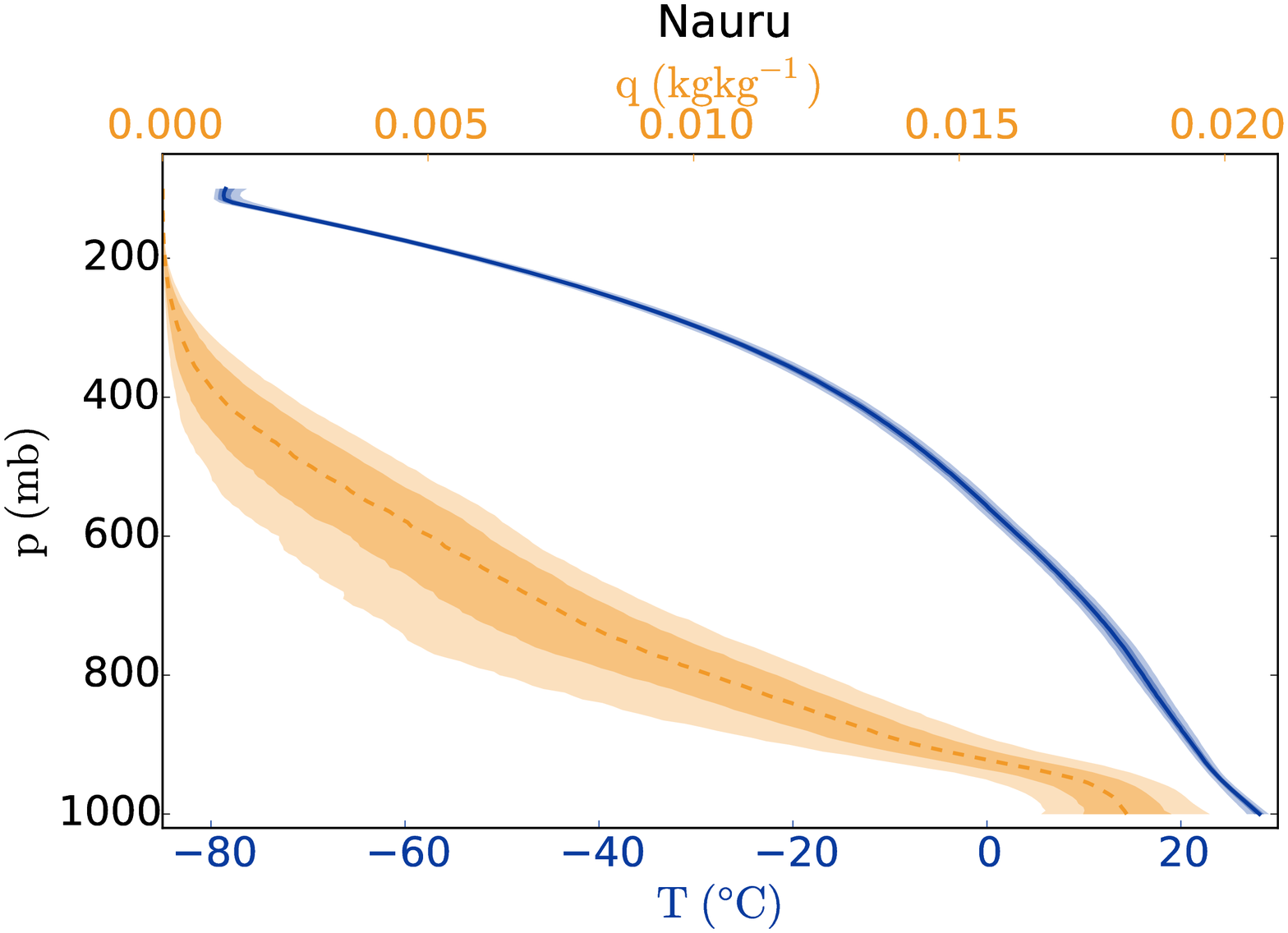}
\caption{Median profiles of temperature $T$ (\degree C, solid line) and specific humidity $q$ (kgkg\textsuperscript{-1}, dashed line) for the ARM soundings used to test the MAPE sorting algorithms. The dark shading shows the 25\textsuperscript{th} to 75\textsuperscript{th} percentile, and the light shading shows the 10\textsuperscript{th} to 90\textsuperscript{th} percentile.}
\label{fig:median_profiles}
\end{figure*}

To calculate the MAPE of a sounding, the sorting algorithms outlined in Section \ref{sec:algorithms} require the input of the temperature, pressure and total specific humidity profiles. The atmospheric profiles used to compare the algorithms are data obtained through soundings from the Atmospheric Radiation Measurement (ARM) Program \citep{stokes1994atmospheric}. We assume that the total specific humidity $q_T$ in the soundings is equal to the specific humidity $q$, i.e. that no liquid water is present in the atmosphere. This widens our choice of data since we do not require liquid water measurements, and is justified since we do not expect large quantities of liquid water to be residing in the atmosphere for long periods of time.

We have used soundings from Nauru dating from 1 April 2001 to 16 August 2006. These soundings contain data that have been interpolated onto 5 mb pressure levels and quality controlled as described by \citet{holloway2009moisture}. We take all soundings with at least 150 valid measurements of temperature and specific humidity, for which the valid measurements span at least the interval from 1000 mb to 100 mb. Any missing temperature or humidity measurements are filled in by linear interpolation. This results in 3130 soundings for which we can use the sorting algorithms to compute the MAPE in the 1000 mb to 100 mb layer using 181 parcels of 5 mb depth.

To verify whether the performance of the algorithms is significantly affected if the soundings are from a different location, we have also used soundings from the ARM Southern Great Plains (SGP) sites during the Intensive Observation Period from 4 June 1997 to 7 July 1997; this dataset is the one used by \citet{tailleux2004seemingly}.
The pressure levels measured in the SGP soundings vary, so we select those soundings that have at least 2000 valid measurements extending from 950 mb to 100 mb, and no more than 50 invalid measurements, resulting in a total of 584 suitable soundings. We linearly interpolate the temperature and humidity data onto 5 mb-spaced pressure levels between 950 mb and 100 mb (resulting in 171 parcels per sounding), to match the parcel mass of the Nauru soundings. The results of Section \ref{sec:comparison} were found to be insensitive to interpolating to a greater number of parcels. 

The median profiles of temperature $T$ and specific humidity $q$ are shown for each location in Figure \ref{fig:median_profiles}, along with the 25\textsuperscript{th} to 75\textsuperscript{th} percentiles (dark shading) and 10\textsuperscript{th} to 90\textsuperscript{th} percentiles (light shading). The profiles are similar in the two locations, with Nauru soundings exhibiting higher moisture at lower levels (this is reasonable because we have kept Nauru data at higher pressure levels, whereas there were insufficient measurements to do so for the SGP data). The Nauru soundings also show colder temperatures at high altitude. It is notable that there is very little variation about the median Nauru temperature profile, and therefore differences in the ability of the algorithms to accurately calculate MAPE here will be mostly due to the differences in humidity profiles between the soundings.

\section{Comparison of Algorithms}
\label{sec:comparison}
All the sorting algorithms discussed in Section \ref{sec:algorithms} were used to calculate the MAPE of each of the 3714 ARM soundings described in Section \ref{sec:data}. To summarise, these algorithms are: Munkres, Lorenz, Randall and Wang, top-down, bottom-up, divide-and-conquer, and Emanuel. We use \citeauthor{stansifer2017accurate}'s implementation of the Munkres algorithm, modified to speed up computation using iterative methods as described in Appendix \ref{sec:bisection}. For the other algorithms we use our own implementations, incorporating the iterative method approach. The MAPE found by the Munkres algorithm is the maximum possible computable MAPE; in the following section we compare this to the MAPE computed by the approximate algorithms to assess their accuracy.

To quantify the accuracy of each algorithm we define the percentage relative difference in MAPE as
\begin{equation}
\label{eq:DR}
D_R =
\begin{cases}
    \frac{\left|\mathrm{MAPE_{munk}} - \mathrm{MAPE_{app}}\right|}{\mathrm{MAPE_{munk}}+\mathrm{MAPE_{app}}}\times 100,& \text{if } \mathrm{MAPE_{app}}\geq 0\\
    100,              & \text{otherwise},
\end{cases}
\end{equation}
where $\mathrm{MAPE_{app}}$ is the MAPE computed by the approximate algorithm, and $\mathrm{MAPE_{munk}}$ is the MAPE computed by the Munkres algorithm. This provides a measure of the amount of MAPE that each approximate algorithm fails to capture. All the approximate algorithms that are based on sorting parcels must compute a MAPE lower than the value computed by the Munkres algorithm, while the Emanuel MAPE may exceed this value.

\begin{figure*}
\centering
\includegraphics[width=0.45\textwidth]{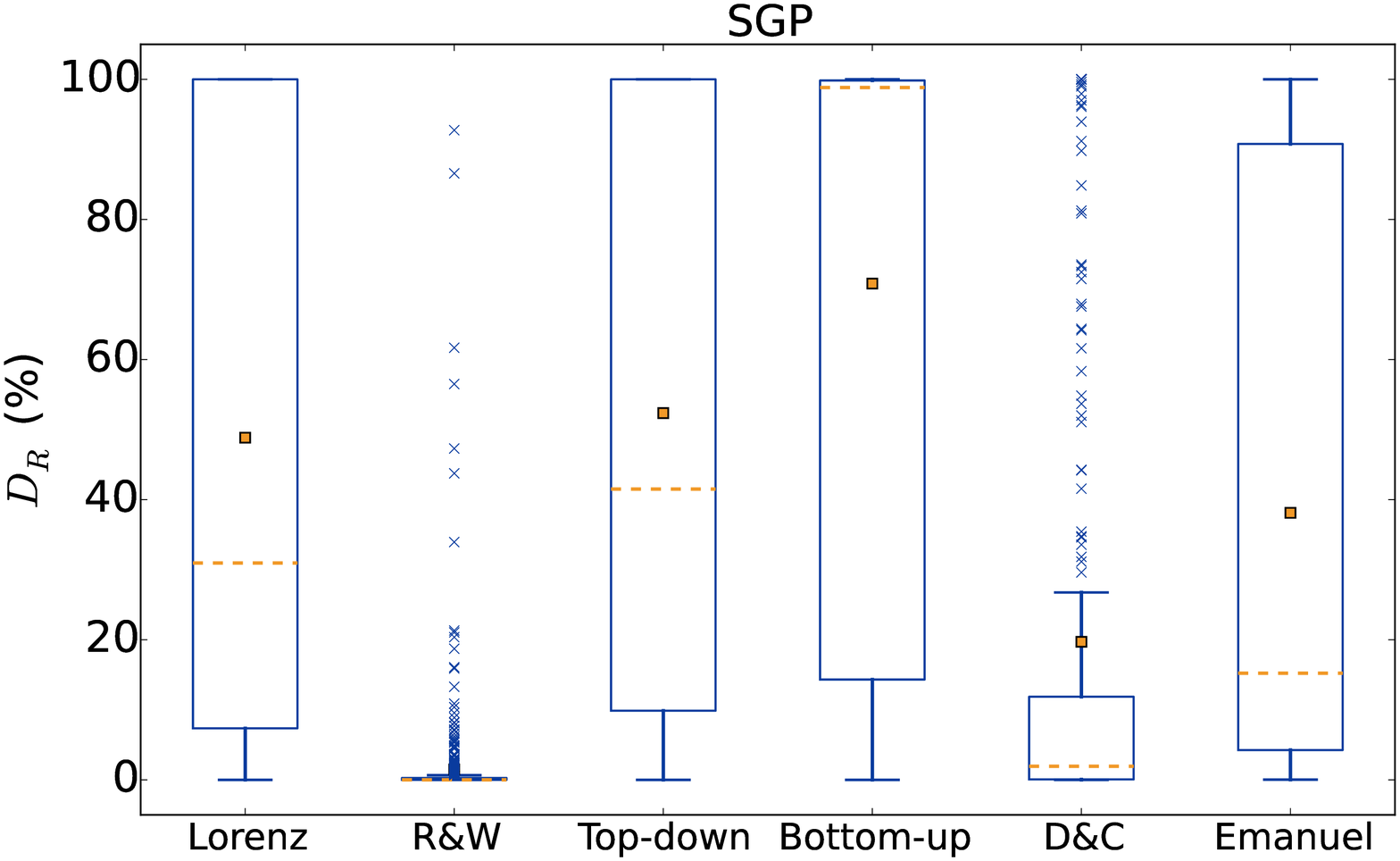}\hspace{3mm}
\includegraphics[width=0.45\textwidth]{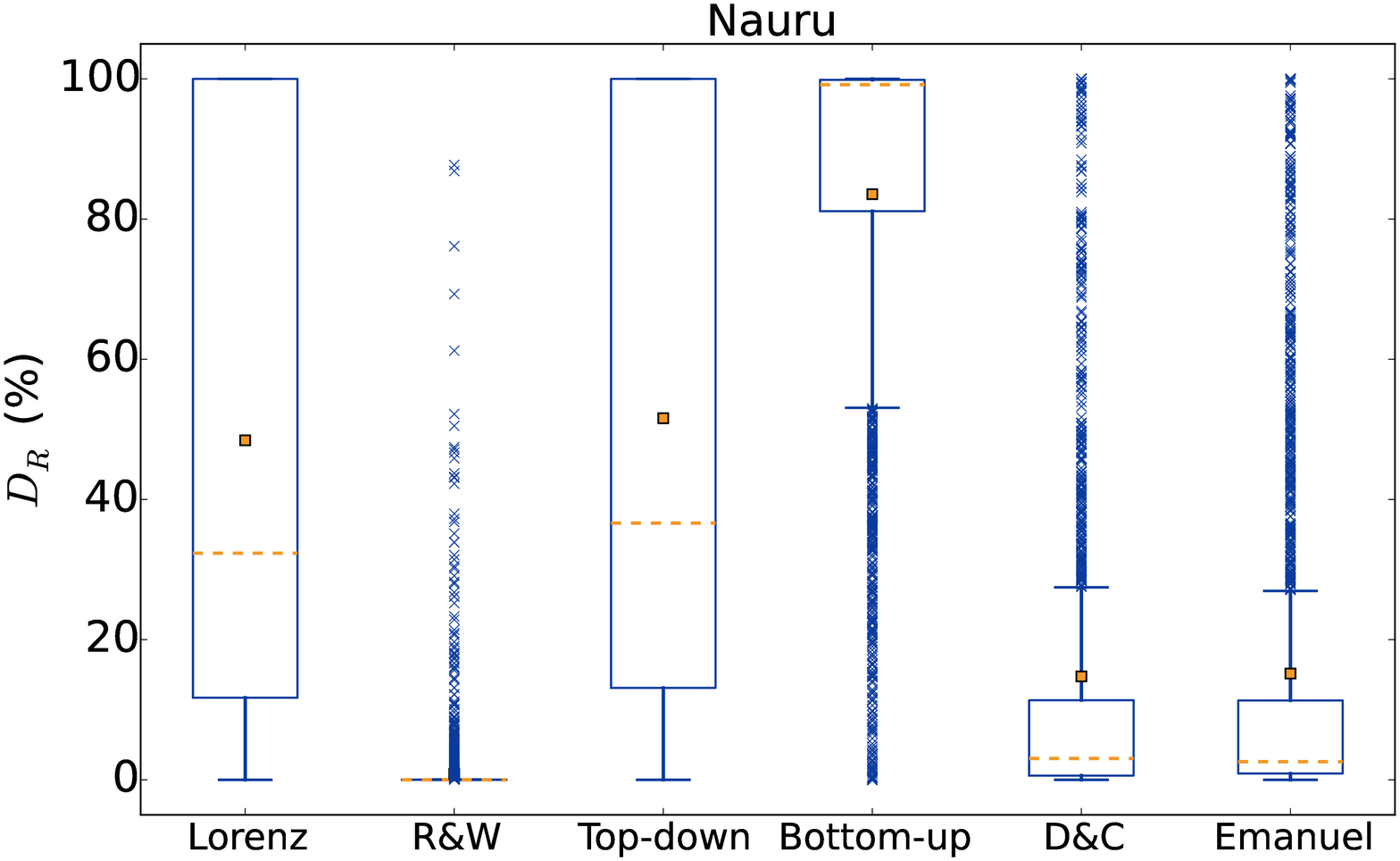}
\caption{Box plots of the percentage relative difference $D_R$, as defined in Eq. (\ref{eq:DR}), between each approximate algorithm and the exact Munkres algorithm. The dashed lines denote the median $D_R$ across the soundings, the squares the mean $D_R$, the boxes the 25\textsuperscript{th} to 75\textsuperscript{th} percentiles, and the whiskers the 10\textsuperscript{th} to 90\textsuperscript{th} percentiles. Crosses represent soundings with outlying $D_R$ values.}
\label{fig:boxplots}
\end{figure*}

The distributions of $D_R$ for each approximate algorithm across all the soundings are displayed in Figure \ref{fig:boxplots}. It is clear from these results that Randall and Wang's algorithm is the most accurate of the six approximate algorithms, with a median $D_R$ of 0.0077\% for the SGP soundings and 0.0015\% for the Nauru soundings. However, there remain outlying cases in which even Randall and Wang's algorithm fails to capture a large proportion of the MAPE. Of the other algorithms, only divide-and-conquer provides a reasonable approximation to the Munkres algorithm, with a median $D_R$ of 1.9\% across the soundings from the SGP, and 3.0\% across those from Nauru.

The bottom-up algorithm fails to capture the majority of the MAPE in most cases; this is expected since the sorting procedure is designed to limit the release of CAPE from buoyant surface parcels, and hence should result in a smaller vertical component of MAPE. There is still a wide range of $D_R$ across the soundings, particularly in the SGP case.
 
\begin{figure}
\centering
\includegraphics[width=0.45\textwidth]{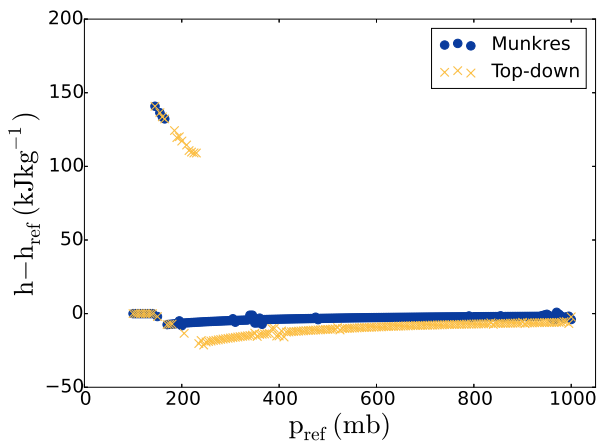}
\caption{Difference between the enthalpy of each parcel from the 24 September 2001 1200 UTC Nauru sounding and its enthalpy in the reference state, plotted against the parcel's pressure in the reference state. Reference states are calculated both using the Munkres algorithm, which computes the exact MAPE, and the top-down algorithm, which computes a negative MAPE.}
\label{fig:td_neg}
\end{figure}

Lorenz's algorithm exhibits a very similar $D_R$ distribution to the top-down algorithm, with both medians around 50\% for each location. The poor accuracy is largely due to the fact that both the top-down and Lorenz procedures frequently compute a negative MAPE, which is unphysical by the definition of MAPE as the difference between the enthalpies of the atmosphere and its rearranged, minimised total enthalpy state. A negative value of MAPE simply means that the ``minimised" enthalpy reference state computed by the approximate algorithm in fact has a higher enthalpy than the real atmospheric state. The top-down and Lorenz algorithms compute a negative MAPE for between 30 to 40\% of the soundings, for both the SGP and Nauru data. To illustrate why this occurs, Figure \ref{fig:td_neg} shows the difference in enthalpy for each parcel between the original sounding and the reference state, for the Nauru sounding measured at 1200 UTC on 24 September 2001. Referring to Eq. (\ref{eq:APE}), the total MAPE of the sounding will be equal to the sum of this enthalpy difference $h-h_{\mathrm{ref}}$ over all parcels. The circular markers show the enthalpy difference when using the Munkres algorithm, which computes a MAPE of 6.87 Jkg\textsuperscript{-1}. The crosses show the enthalpy difference using top-down sorting, which results in a MAPE of -19.9 Jkg\textsuperscript{-1}. It is evident that the negative MAPE is a result of the top-down algorithm lifting parcels to low reference pressures (150-250 mb), in such a way that these particular parcels experience a large decrease in enthalpy, but the parcels at higher reference pressures exhibit a slight increase in enthalpy. When computing the overall MAPE, the smaller enthalpy decreases over many parcels at high reference pressures outweigh the large enthalpy decreases of the few parcels at low reference pressures. Parcels at high reference pressures in the Munkres reference state also show a decrease in enthalpy, but, since this decrease is smaller than in the top-down case, the net MAPE remains positive. The divide-and-conquer algorithm also computes negative MAPE for some soundings, but this does not occur as frequently as for the Lorenz and top-down algorithms (13\% of SGP soundings, 6.5\% of Nauru soundings). The bottom-up, Randall and Wang, and Emanuel algorithms do not compute a negative MAPE for any sounding.

\begin{figure*}
\centering
\includegraphics[width=0.45\textwidth]{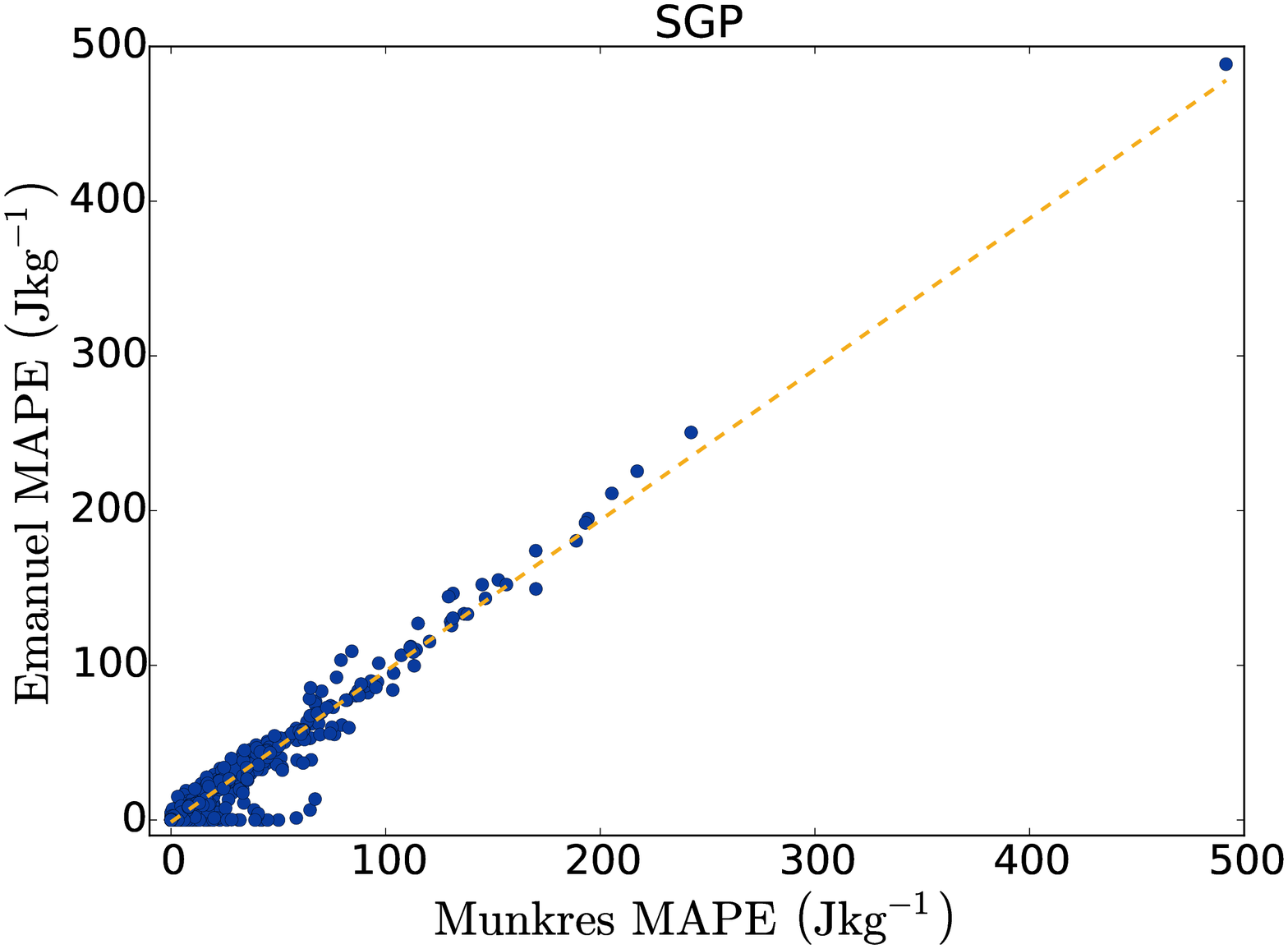}\hspace{3mm}
\includegraphics[width=0.45\textwidth]{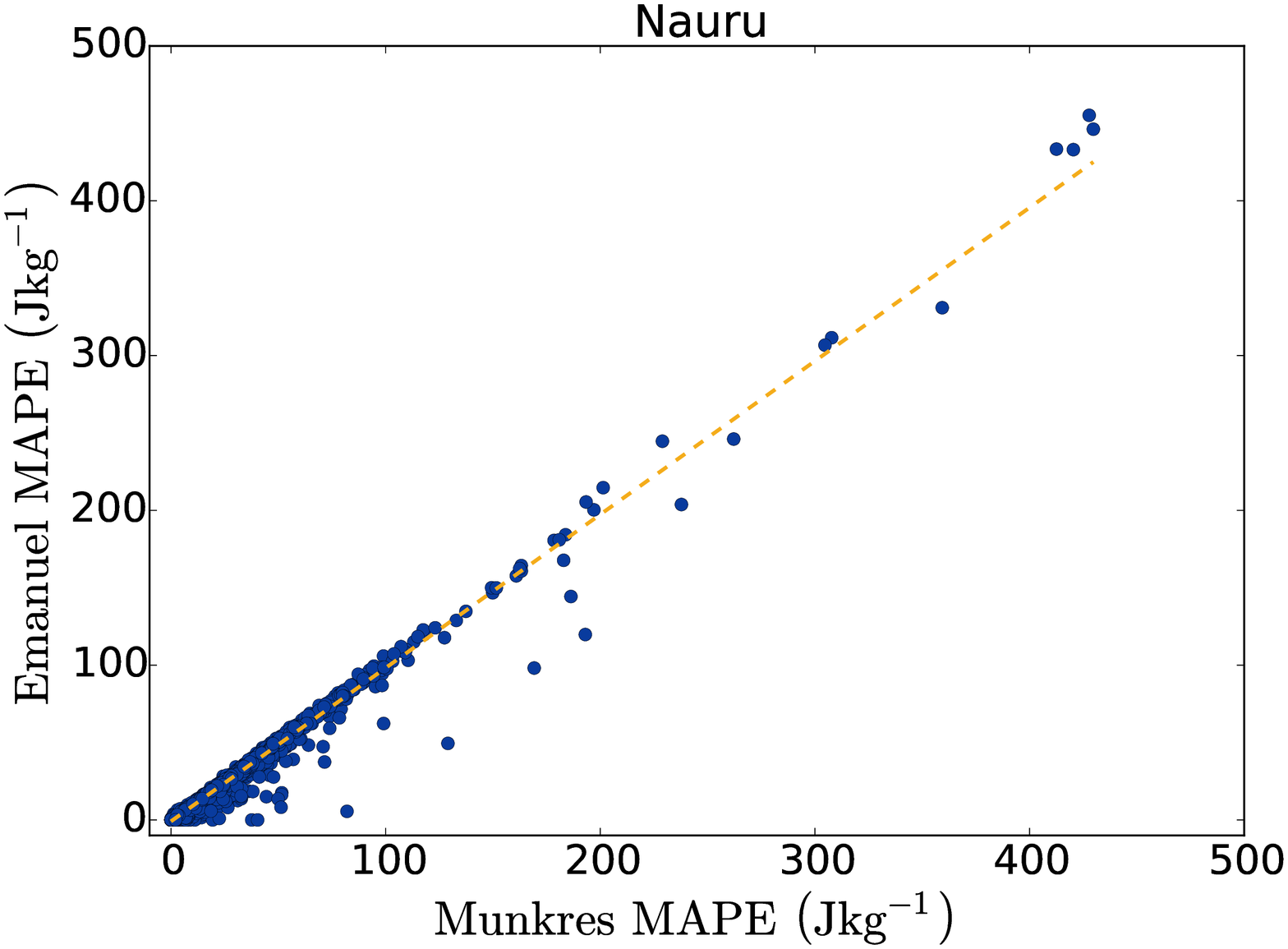}
\caption{Comparison of MAPE calculated using the Emanuel algorithm against the exact MAPE calculated using the Munkres algorithm. The dashed line in each case shows the linear best fit to the data.}
\label{fig:emanuel_scatter}
\end{figure*}

In general the $D_R$ distributions of the sorting-based algorithms are similar for the two locations, suggesting that we do not expect the optimum choice of sorting algorithm to change depending on the typical local atmospheric conditions. However, the accuracy of the Emanuel MAPE is very different between the locations. For the Nauru soundings, it has a median $D_R$ of $2.6\%$, which is comparable to the divide-and-conquer sorting algorithm, showing that the Emanuel algorithm would be a sensible choice for estimating MAPE. In contrast, its median $D_R$ over the SGP is $15\%$, which is much less accurate than either the divide-and-conquer or Randall and Wang algorithms, and so the Emanuel method would not be a good practical choice for computing MAPE in this environment. To investigate why this difference in accuracy occurs, Figure \ref{fig:emanuel_scatter} shows scatter plots of the Emanuel MAPE against the MAPE computed using the Munkres algorithm, for each location. The red dashed lines display the best linear fit to the data; for the Nauru soundings we find a correlation coefficient of $r = 0.990$, while for the SGP soundings the correlation coefficient is slightly poorer, at $r = 0.977$, as expected from the higher median value of $D_R$. The high correlation indicates that most of the MAPE present in the ARM soundings corresponds to the CAPE of near-surface parcels, in line with the assumption of \citet{emanuel1994atmospheric}. We can see from Figure \ref{fig:emanuel_scatter} that the poorer correlation for the SGP compared to the Nauru soundings is mostly due to a number of SGP soundings that have very low Emanuel MAPE, but values of Munkres MAPE up to 60 Jkg\textsuperscript{-1}. We find that these discrepancies arise where an unstable layer that is elevated from the surface is present in the sounding. Since the Emanuel algorithm we have used assumes that the CAPE-containing boundary layer begins at the surface, these elevated instabilities are not correctly captured. This issue could be solved by designing an algorithm that varied both $\Delta p_b$ and the pressure of the boundary layer bottom, although this would increase the computational expense. The main advantage of the Emanuel algorithm is that it provides greater physical insight into how MAPE can be converted to kinetic energy via convection, rather than relying on the physically unconstrained rearrangements of a sorting algorithm, as will be discussed further in Section \ref{sec:discussion}.

\section{Discussion}
\label{sec:discussion}
The results presented in Section \ref{sec:comparison} allow us to make a more informed assessment of which algorithms are most suitable for the computation of MAPE, based on an analysis of a wider range of soundings than in previous studies.  The key challenge for MAPE algorithms stems from the fact that MAPE is ultimately a residual between the positive work due to the release of CAPE minus the negative work due to compensating subsidence. As a result, sorting a vertical sounding according to decreasing density, which is the approach underlying the majority of algorithms, may occasionally result in a reference state with a larger potential energy than the actual state, if the negative work exceeds the positive work. This is in contrast to the case of a dry atmosphere, for which sorting the actual state according to potential temperature always returns the state of minimum potential energy. Without an explicit procedure to forbid it, most heuristics for computing MAPE are bound to return a negative value in some cases. That such situations may occur in practice appears to have been overlooked in previous studies, with the exception of \citet{randall1992moist}, but is clearly established for the particular soundings analysed here.

In terms of performance, we have found that the Lorenz and top-down algorithms have nearly identical levels of accuracy, which we did not anticipate. However, our results also indicate that both algorithms are so prone to returning a negative MAPE that they are not suited to practical application. Such an issue was not mentioned in previous studies using these algorithms \citep{ogorman2010understanding,wong2016computation}, possibly because only positive MAPE values were found in the specific cases analysed.

The algorithm introduced by \citet{randall1992moist} was found to be a good predictor of the exact MAPE across the soundings studied. The relative difference between the Randall and Wang and Munkres algorithms exceeded 10\% for only 2\% of the soundings studied. This algorithm also benefits from the fact it is specifically designed to never return a negative MAPE. However, it is the most computationally expensive of the approximate algorithms; for a sounding with a small number of parcels it takes even longer to run than the Munkres algorithm (see Appendix \ref{sec:bisection}).

As was outlined in Section \ref{sec:comparison}, the divide-and-conquer algorithm is the only other approximate sorting method showing reasonable accuracy over the soundings studied here. It is also by far the fastest of the approximate algorithms. We therefore conclude that the divide-and-conquer algorithm is the best option for the approximation of MAPE, since it offers a balance between accuracy and speed, as was suggested by \citeauthor{stansifer2017accurate}. On the other hand, the relative difference between the divide-and-conquer and Munkres algorithms is greater than 50\% for 11\% of the total soundings studied, and divide-and-conquer sorting may result in a negative MAPE. This clearly suggests that the three test cases analysed by \citeauthor{stansifer2017accurate}, for which the algorithm was found to perform well, might be special cases which are not sufficiently representative of the variety of situations that can be encountered in nature.

If the exact MAPE is required, then the Munkres algorithm is the only choice; in that case, we recommend the use of the bisection method outlined in Appendix \ref{sec:bisection}, as it considerably increases the computational efficiency of the method.

In Section \ref{sec:comparison}, we also demonstrated the feasibility and good performance of an algorithm exploiting \citet{emanuel1994atmospheric}'s theoretical expression for MAPE, which does not involve any form of sorting nor require discretising the vertical sounding into parcels of equal mass, and which by construction always return a positive value of MAPE.

So far, the implicit assumption of the present study and others has been that it is legitimate or most useful to define the APE of a moist atmosphere in terms of the reference state that defines the absolute minimum in potential energy, but this is not necessarily the case. For a moist atmosphere, it is a priori possible to construct alternative sorted reference states that define only a local minimum in potential energy. Although such reference states would result in a lesser global value of APE, it is unclear why this would necessarily invalidate their use. In tropical cyclones, for instance, numerical simulations reveal that boundary layer parcels away from the eyewall may have CAPE whose release is suppressed by the subsidence in that region, as pointed out by \cite{wong2016computation}. Since the CAPE of such parcels can rarely if ever be released, it is unclear why it should be included in the definition of a tropical cyclone APE, as will normally be the case if the reference state defining a global potential energy minimum is selected. From a practical viewpoint, it is important to remark that the choice of reference state affects the overall value of APE as well as its diabatic generation rate $G(APE)$, but neither affects the energy conversion between APE and kinetic energy, nor the general form of the APE evolution equation, given by:
\begin{equation}
     \frac{{\rm d}APE_i}{{\rm d}t} = C(KE,APE) + G(APE)_i ,
     \label{eq:APE_evolution_equation}
\end{equation}
where the index $i$ is used to indicate dependence on the reference state chosen. Eq. (\ref{eq:APE_evolution_equation}) states that the conversion $C(KE,APE)$ between kinetic energy and APE always appears as a residual between the APE storage term ${\rm d}APE_i/{\rm d}t$ and the APE generation rate $G(APE)_i$ (see \citet{pauluis2007sources} for a discussion of how moist processes may affect the latter). From a theoretical viewpoint, Eq. (\ref{eq:APE_evolution_equation}) represents a balance between three terms, of which the storage term is the least interesting or meaningful. For this reason, \citet{wong2016computation} argued that the reference state should be chosen so as to minimise the storage term, in order to potentially make it possible to predict the APE/KE conversion from the knowledge of the APE generation rate. In this regard, \citet{wong2016computation} found the use of the bottom-up sorted reference state to yield a lower storage term that the top-down sorted reference state, but more research is required to establish whether this can be regarded as a general result. 

Given the computational and conceptual difficulties entailing their use, it is important to question whether sorting algorithms are really needed to study the energetics of a moist atmosphere. The idea that simpler alternatives might exist is indeed justified by the fact that some recent APE studies successfully moved away from the use of sorting algorithms by resorting to p.d.f. approaches instead, as in the case of \citet{saenz2015estimating}, itself an extension of \citet{tseng2001mixing}, although it is unclear how such a method could be applied to a moist atmosphere. Also, it has long been known from the works of \citet{andrews1981note} and \citet{holliday1981potential} that it is possible to construct a local theory of APE based on an arbitrary reference state defined by a reference pressure $p_0(z,t)$ and specific volume $\alpha_0(z,t)$ in hydrostatic equilibrium. Based on \citet{tailleux2013available} and \citet{novak2017local}, this would lead one to define the APE density for a moist atmosphere as the work that a fluid parcel needs to perform to move from its reference pressure $p_r$ to its actual pressure $p$, viz., 
\begin{equation}
    e_a(\theta_l,q_T,p,t) = \int_{p_r}^p \left [ \alpha(\theta_l,q_T,p') - \alpha_0(p',t) \right ] {\rm d}p' ,
    \label{APE_density}
\end{equation}
where $\theta_l$ is liquid potential temperature and $q_T$ is total water content. An alternative formulation for APE density in a compressible atmosphere, based on a modified potential temperature, has been proposed by \citet{peng2015local}. In contrast to what is often assumed, a sorting algorithm is not required to calculate the reference pressure $p_r$. Indeed, as shown by \citet{tailleux2013available}, if $\alpha_0(p,t)$ is known at all times as a function of pressure, $p_r$ can be simply estimated by solving the so-called Level of Neutral Buoyancy (LNB) equation:
\begin{equation}
      \alpha(\theta_l,q_T,p_r) = \alpha_0(p_r,t) .
      \label{LNB_equation}
\end{equation}
This corresponds to the use of a level of neutral buoyancy in the Emanuel MAPE algorithm, demonstrating the link between the local and global approaches to APE.

Illustrations of how to construct energy budgets in the oceans and dry atmosphere in the case where the reference density profile is defined from a horizontal or isobaric average are discussed \citet{tailleux2013available} and \citet{novak2017local} respectively. These recent developments, combined with the physical insights brought about by \citet{emanuel1994atmospheric}'s theoretical expression for MAPE, suggest that a satisfactory theory of available potential energy for a moist atmosphere, which has been lacking so far, might be at hand provided that one moves away from sorting algorithms altogether, as we hope to demonstrate in subsequent studies.

\acks 

BLH is funded by NERC as part of the SCENARIO Doctoral Training Partnership (NE/L002566/1). We would like to thank Chris Holloway and Pier Luigi Vidale for interesting discussions, and Eric Stansifer for making the code to compute MAPE using the Munkres algorithm available. All data and algorithms used to obtain the results presented in this paper are available upon request (b.l.harris@pgr.reading.ac.uk).

\appendix
\section{Increasing Algorithm Efficiency}
\label{sec:bisection}
All the algorithms used to compute MAPE in this work require the repeated calculation of the temperature of an air parcel when it moves reversibly adiabatically to a given pressure level. For the approximate sorting algorithms, this calculation is required in order to sort the parcels by density at each pressure level of the sorting process. The Emanuel algorithm requires the calculation in order to find the specific volume of lifted parcels for the calculation of CAPE (see Eq. (\ref{eq:CAPE})).  For the Munkres algorithm, it is required in order to compute the \textit{cost matrix} $\mathbf{C}$, where $c_{ij}$ is the enthalpy of the $i$\textsuperscript{th} parcel if it were moved to the $j$\textsuperscript{th} pressure level. This cost matrix is then manipulated to find the minimum enthalpy rearrangement of the parcels. Such temperature calculations are a time-consuming stage of the algorithms because they require the use of a root-finding procedure. In this appendix we show how the bisection method can be employed to enable vectorised calculation of the new temperatures, so that the runtime of the algorithms is reduced by moving many parcels to new pressure levels at once.

The problem can be formulated as follows: we know the initial temperature $T$ (\degree C), pressure $p$ (mb) and total specific humidity $q_T$ (kgkg\textsuperscript{-1}) of the parcel, and the target pressure $p'$. We must find $T^*$, the temperature such that $\Delta s = s(T,p,q_T) - s(T^*,p',q_T) = 0$, where $s$ (Jkg\textsuperscript{-1}K\textsuperscript{-1}) is the specific entropy of the parcel (so that the movement of the parcel is adiabatic). Using a built-in root-finding function in Python  (such as Brent's method) to obtain the value of $T^*$, as has been done previously, necessitates looping through all parcels individually to calculate their temperatures, since built-in root-finding functions can compute the root of only one function at a time.

Alternatively, it is possible to use a simple bisection method to compute $T^*$. This converges less quickly than other root-finding procedures (its convergence is linear), but makes it easy to compute $T^*$ for many parcels at once. We initially take a wide temperature interval from 0 K to 1000 K. This interval is successively halved, at each stage identifying which half $T^*$ lies in by computing the sign of $\Delta s$ at the midpoint of the interval. The bisection method is guaranteed to converge to the root $T^*$ provided that the function $\Delta s(T)$ is monotonic, $\Delta s\left(0\;\mathrm{K}\right)<0$, and $\Delta s\left(1000\;\mathrm{K}\right)>0$. The number of iterations can be adapted depending on the required accuracy of $T^*$; we have ensured convergence to within $10^{-6}$K.

\begin{figure}
\centering
\includegraphics[width=0.45\textwidth]{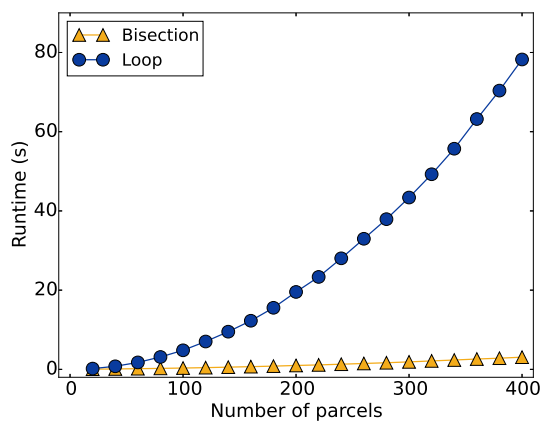}
\caption{Time taken for the Munkres algorithm to compute the MAPE of an $n$-parcel sounding linearly interpolated from the data at Nauru on 1 April 2001, 1200 UTC, for the algorithm with and without the bisection method incorporated.}
\label{fig:munkres_time}
\end{figure}

To assess the impact of using the bisection method in this way, we have created two versions of the Munkres algorithm provided by \citet{stansifer2017accurate}: one that calculates the cost matrix by looping through all parcels and using Brent's method to compute temperature changes (similar to \citeauthor{stansifer2017accurate}'s original algorithm); and one that uses the bisection method to compute one row of the cost matrix at a time. We use both these algorithms to compute the MAPE for soundings with a varying number of parcels. For each number of parcels $n$ we create the desired sounding by taking the ARM Nauru data from 1200 UTC on 1 April 2001 and linearly interpolating it to $n$ pressure levels. The time taken to compute the MAPE on a personal computer is displayed in Figure \ref{fig:munkres_time}; in all cases it was verified that the two algorithms computed the same MAPE. It is clear that use of the bisection method results in a much faster algorithm for large numbers of parcels. 

\begin{figure}
\centering
\includegraphics[width=0.45\textwidth]{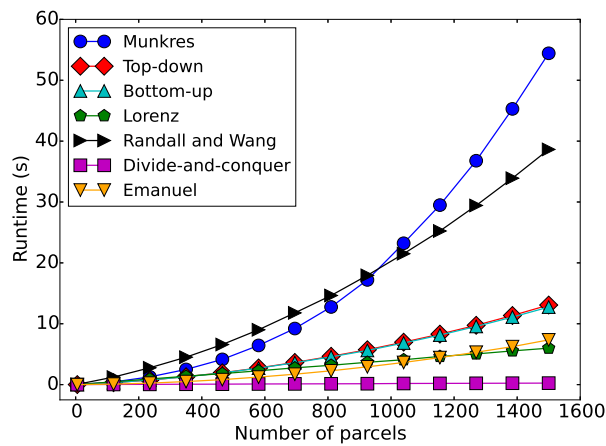}
\caption{Time taken for each MAPE algorithm to compute the MAPE of an $n$-parcel sounding linearly interpolated from the data at Nauru on 1 April 2001, 1200 UTC.}
\label{fig:runtimes}
\end{figure}

The bisection method can be similarly employed for the approximate sorting algorithms. Figure \ref{fig:runtimes} shows the time taken to compute MAPE for our implementations of each of the algorithms described in the main body of the paper, for a varying number of parcels. As expected, the exact Munkres procedure is slowest for large numbers of parcels. The divide-and-conquer algorithm of \citet{stansifer2017accurate} is easily the fastest, making it a good compromise between speed and accuracy.

\bibliographystyle{wileyqj}
\bibliography{ref.bib}

\end{document}